\documentstyle[12pt ]{article}
\input epsf
\begin{document}

\begin{titlepage}
\begin{flushright}
IC/96/239
\end{flushright}
\medskip

\begin{center}
{\Large\bf SU(7) SUSY GUTs \\
  with a natural intermediate scale} \\
\vspace{7mm}
{\bf Ilia~~Gogoladze  \footnote {E-mail address: Gogoladz@mail.cern.ch}}
\vspace{4mm}

{\it Institute of Physics of Georgian Academy of Sciences,
Tamarashvili 6,\\
Tbilisi 380077,  Georgia}  \\
\medskip
{\it and}\\
\medskip
{\it International Centre for Theoretical Physics, Trieste,Italy}

\end{center}
\vspace*{3cm}

\begin{abstract}

We investigate the  $SU(N)$
supersymmetric  Grand Unified Theories  with ``custodial symmetry''
mechanism to  explane  the doublet-triplet
hierarchy.
We show that in such type of $SU(7)$ SUSY theory intermediate scale
appears naturally and 
the correct value for $sin^2 {\theta}_{W}$ is predicted via 
vector-like matter superfields splitting. The 
unification appears to be closed to $M_{Pl}$ for all the reasonable
values of
$\alpha_{s}$ and $M_{SUSY}$.
Due to the large unification scale the baryon number violating $d=5$
operator is suppressed in comparison with that in  minimal $SU(5)$
theory.

\end{abstract}

\end{titlepage}

\section{Introduction}

Perhaps the  gauge hierarchy problem is main difficulty of the  Grand
Unified Theories  
 (GUT). This problem can be stated
 into the following  questions: a) why is the electro-weak scale
stable against the radiative corrections? b) how do the Higgs doublets
remain light, whereas their colour triplet partners  must be superheavy
in order to avoid fast proton decay?

 As it is well-known [1] supersymmetry (SUSY) answers the first question,
if its breaking scale is $M_{SUSY}\sim 1$ TeV. This is one of 
 the main motivations for            
 the low energy  SUSY.
   
Several attempts to
answer the second question (which is known as DT splitting problem) were
suggested in the literature. Possible solution to
the  DT splitting problem to be explored below 
could be
due ``custodial symmetry'' mechanism. [2]  (which is discussed 
Sect. 2.)

 It is  well known   that the combined
analysis, including the heavy and light threshold corrections, in the
minimal $SU(5)$ GUT predicts a value of $\alpha_{s}\approx 0.126$[3]
when the superpartners masses are at the
TeV scale.  If the sparticle masses are at the $500$ GeV scale the
predicted value of $\alpha_{s}$ is $\approx 0.13$. On the other hand it is
known that
from $Z$-peak we have $\alpha_{s}=0.118^{+0.004}_{-0.007}(exp)\pm 
0.002(theor)$ [4]

 So to reduce
the predictions of $\alpha_{s}$ in the SUSY GUT one needs a high value of
$M_{SUSY}$ which,on the other hand, is unnatural for the stability of the 
electro-weak scale.

Among the existing attempts to solve this problem we quote one the $SU(5)$
SUSY theory with missing
partner mechanism with   a scalar content  $75$; $50$ and $\overline
{50}$ representation. As was shown [3] the heavy threshold effects
coming from those multiplets at the GUT scale make possible to get
a low value of $\alpha_{s}$ when $M_{SUSY}\leq$ 1 TeV.

Another attempt was proposed by Brahmachari and
Mohapatra [5] in the SUSY $SO(10)$  theory with intermediate gauge
$(G_I= SU(3)\otimes SU(2)_L \otimes SU(2)_R \otimes U(1)_{B-L})$
symmetry at scale ${10}^{10}\div {10}^{12} ~GeV$. 

In the present paper we propose an alternative scenario: 
in our case we have $SU(3)_{C}\otimes SU(3)_{W}\otimes U_1(1)$ intermediate 
scale which is independently motivated from the natural solution 
of the doublet-triplet splitting problem, and at the same time 
is an outcome of the theory in terms of the GUT and the week 
(or low energy SUSY breaking) scale. 
Using split-multiplet mechanism is possible to make unification for
all the reasonable values of $\alpha_s$ and $M_{SUSY}$.  
\section{DT splitting mechanism}

Consider a $SUSY$ GUT based on $SU(6)$ gauge symmetry
with a minimal set of  Higgs superfields needed for  the breaking of
$SU(6)$
symmetry down to SM. These are adjoint 35-plet ($\Sigma$) and a pair of
fundamental $(6+\overline 6)$-plets $(H+\overline H)$. The most general
$SU(6)$ invariant renormalizable scalar superpotential of the Higgs 
superfields has the following form:
\begin{equation}
W=\frac{m}{2}Tr\Sigma^{2}+ \frac
{h}{3}Tr\Sigma^{3}+\lambda\overline{H}\Sigma H+M\overline{H} H
\end{equation}
where $h$ and $\lambda$ are the dimensionless constants, $m$ and $M$
are mass parameters (we suppose that M $\gg$ m). One of the possible VEV
configuration of this fields (in unbroken SUSY limit) is:
\begin{eqnarray}
< \Sigma >=diag[1~ 1~ 1~ -1~ -1~ -1]\frac{M}{\lambda}+diag[2~ 2~ 2~ -3~ -3~0]
\frac{m}{h}    \nonumber \\
    \nonumber                              \\
< \overline H > = < H > = [0 ~0 ~0 ~0 ~0 ~1]
\biggl( 6\frac{m}{\lambda}\biggl(\frac{M}{\lambda}+\frac{m}{h}\biggl)\biggl)^{1/2}
\end{eqnarray}

Obviously, with this solution the hierarchy of the
symmetry breakings is the following: at the scale $\sim M$ the gauge
symmetry is broken down to $SU(3)_{C}\otimes SU(3)_{W}\otimes U_1(1)$ by
the $35$-plet which develops a large VEV; then at the  ``geometrical
scale'' $\sim \frac{\sqrt{Mm}}{\lambda}$ the VEVs
$<H>=<\overline H>$  break
$SU(3)_{W}\otimes U_{1}(1)$ to  $SU(2)_{W}\otimes U(1)_{Y}$.
Obviously, in this case we have no light Higgs doublets.

Now let us  consider an extension of the $SU(6)$ symmetry to
$SU(6)\otimes SU(N)_{cust}$ [2], 
where we introduce $N 6, \overline{6}$ - plet superfields 
$H_A \overline{H^A} \, 
(A=1,2...N)$ transforming as $N,\overline{N}$ representations of
$SU(N)_{cust}$. We have the following superpotential
\begin{equation}
W~=~\frac{m}{2}Tr\Sigma^{2}~ + ~\frac
{h}{3}Tr\Sigma^{3}~+~\lambda\overline{H}^{A}\Sigma
H_{A}~+~M\overline{H}^A ~H_{A}
\end{equation}

and we  choose the solution for which a single pair  $(\overline
H^A+H_A)$
 (say for A = 1) develops a  VEV. The VEVs of $\Sigma$ and $(\overline
H^{1}+H_{1})$ have the (2) form  and relevant mass matrix for
doublet (antidoublet) fragments from $\Sigma$ and $(\overline{H}^A+H_A)$
is:

\begin{eqnarray}
\overline D_{\Sigma}~ \overline D_{\overline H^{1}}
~\overline D_{\overline H^{2}}~\cdot~ \cdot~
\overline D_{\overline H^{N}}~  \nonumber \\
   \nonumber        \\
\begin{array}{l}
D_{\Sigma} \\
D_{H_{1}} \\
D_{H_{2}} \\
~~\cdot     \\
~~\cdot     \\
D_{H_{N}} \\
\end{array}
~\left [ \begin{array}{cccccc}
a &\sqrt{\frac{ab}{2}} & 0 & \cdot & \cdot & 0 \\
\sqrt{\frac{ab}{2}} & b & 0 & \cdot & \cdot & 0 \\
0 & 0 & b & \cdot & \cdot & 0   \\
\cdot & \cdot & \cdot & \cdot & \cdot & \cdot   \\
\cdot & \cdot & \cdot & \cdot & \cdot & \cdot    \\
0 & 0 & 0 & \cdot & \cdot & b  \\
\end{array} \right ]
\end{eqnarray}
where $D (\overline D)$
stands for a doublet (antidoublet) from the suitable representation
 and
\begin{eqnarray}
a=-~2 \frac{h}{\lambda}M+2m \nonumber   \\
                                \nonumber   \\
b=-~3 \frac{\lambda}{h}m
\end{eqnarray}

    The zero mass eigenstates of the above mass matrix
\begin{center}
$G=[D_{\Sigma}~ \sqrt{b}+D_{H_{1}}~\sqrt{a}~][b+a]^{-1/2}$ \\
$\overline{G} = [\overline{D_{\Sigma}} \sqrt {b}+
\overline{D_{\overline{H_1}}} \sqrt{a}~][b+a]^{-1/2}$
\end{center}

 are eaten up Goldstone superfields. The orthogonal superpositions get
the large masses $\sim M$. As it is clear from eq.(4) the physical Higgs
doublet from
the $({\overline{H}_A}+H^{A}) ~(A\not= 1)$ automatically acquire small
masses ($b=3 \frac{\lambda}{h} m$).

So, we get $(N-1)$ pairs of the  physical Higgs doublets with masses 
$\sim m$

\section {The model}
As was shown in [6] in the case of $SU(6)\otimes SU(2)_{cus}$-symmetry 
 we have no acceptable unification of the gauge constants. 
From the unification condition we get that ``custodial'' symmetry must be
larger than $SU(2)$. 
On the other hand  the $SU(6)\otimes SU(3)_{cus}$ symmetry 
case exhibits is the problem in the fermion sector, namely for the top
quark mass.
This is because the masses of the ``up'' type quarks are induced from
the following  effective nonrenormalizable  couplings (A,B.C is the
indexes
of $SU(3)_{cus}$ symmetry)
\begin{center} 
  $f \cdot \frac{1}{M_{20}}  \cdot 15 \cdot 15 \cdot
 H^{A} \cdot H^{B} \cdot S^{C} \cdot \epsilon_{ABC} $
\end{center} 

(where $S^{C}$ is singlet under $SU(6)$ symmetry and suppose 
$<S^{C}>\sim M_{GUT}$) that are trough the generated by heavy particle 
exchange [10], transforming as $(20^{A}_{1}+\overline{20_{1A}})$ 
end $(20^{A}_{2}+20_{2A})$
representations. If we assume that $(20^{A}_{1}+\overline{20_{1A}})$ 
end $(20^{A}_{2})+20_{2A})$
multiplets(or one of them) have masses $\sim <H>$, in
this case 
the gauge couplings 
 will enter in the nonperturbative regime before unification point.
Increasing masses of 
$20^{A}_{1}+20_{1A}$ end $(20^{A}_{2}+20_{2A})$ multiplets,say
 $\sim M_{GUT}$, will decrease the top mass unacceptably. We will show
that this problem is solved in 
$SU(7) \otimes SU(3)_{cus}$ symmetry case.

Let us choose the Higgs content of our  $SU(7)\otimes SU(3)_{cust}$
model as : $\Sigma(48,1)+\overline{H}_{A}
(\overline{7}_{A},3)+ H^{A}(7^{A},3)+\Phi(7,1)+\overline\Phi(\overline
 7,1)$.where the brackets are indicated  their transformation properties
under the
$SU(7) \otimes SU(3)_{cust}$ group. We also introduce 3 pairs
of $SU(7)$- triplet superfields
${S_{i}}^{A}+{S^{i}}_{A}$ (i=1,2,3) and assume a $Z_2$-symmetry under
which  
$(\overline \Phi,~\Phi) =~-(\overline \Phi,~\Phi)$

Consider the most general $SU(7)\otimes SU(3) \otimes Z_{2}$-invariant
renormalizable scalar superpotential .  ($SU(7)$
indexes are omitted )

\begin{eqnarray}
W~=~\frac{m_{1}}{2}Tr\Sigma^{2}~+~\frac{h}{3}Tr\Sigma^{3}~+~
\lambda_{1}\overline{H_{A}}\Sigma
~H^{A}~+~M_{1}~\overline{H_{A}}~H^{A}~+
~\lambda_{2}\overline{\Phi}\Sigma
~\Phi ~+ \nonumber \\
~m_{2}\overline{\Phi}\Phi~+~M_{2}~{S_{1}}^{A}{S^{1}}_{A}~+
+~M_{3}~{S_{2}}^{A}{S^{2}}_{A}~+
~M_{4}~{S_{3}}^{A}{S^{3}}_{A}~+ \nonumber \\
~h_{1}{S_{1}}^{A}{S_{2}}^{B}{S_{3}}^{C}~\epsilon _{ABC}~+
~h_{2}{S^{1}}_{A}{S^{2}}_{B}{S^{3}}_{C}~\epsilon ^{ABC}~~~~~~~~
\end{eqnarray}

where $\lambda_{1}$, $\lambda_{2}$,
 $h, ~h_{1}$ and $h_{2}$  are the dimensionless constants,
 $M_{1},  ~M_{2},~M_{3},~M_{4}, ~m_{1}$ and $m_{2}$ are the
mass parameters.  We assume that $m_{1} \sim m_{3/2}$ and
$m_{2}\sim m_{3/2}$, $M_{1}, M_{2}, M_{3}$ and $M_{4}$ are of order of $
M_{GUT}$.  Note that introduction of the small mass term
$m_{1}Tr\Sigma ^{2}$ and $m_{2}(\overline{\Phi}\Phi)$ in (6) is not
necessary, since it could be generated after SUSY breaking.  Namely,
in the simplest version of the minimal $N=1$ supergravity SUSY
violating terms lead to the well known scalar potential [7]

\begin{equation}
V=\biggl|\frac{\partial W}{\partial z_i}+m_{3/2}z^{*}_{i}\biggr |^2+
m_{3/2}(A-3)[W^{*}+W]+D-terms
\end{equation}
It is clear, that if one puts $m_{1}=0$ $m_{2}=0$ in (6) the mass term
$m_{3/2}Tr\Sigma ^{2}$ and $m_{3/2}(\overline{\Phi}\Phi)$ ~will be
automatically  generated from (7).

Among the discretely degenerate SUSY minima of (6) the one 
 of our interest is:

\begin{eqnarray}
<\Sigma >=diag[1~ 1~ 1~ -1~ -1~ -1~
0]\frac{M_{1}}{\lambda_{1}}~+~diag[1~ 1~ 1~ -1~ -1~ 0~ -1]\frac
{m_{2}}{\lambda_{2}}~+   \nonumber  \\
~diag[2~ 2~ 2~ -3~ -3 ~0 ~0]
\frac{m_{1}}{h}~~~~~~~~~~~~~~~~~~~~~~~~~~~~~~~~~~~~ \nonumber \\
\nonumber \\
< \overline H_{A}> = <H^{A} > = \left [
 \begin{tabular}{ccccccc} 0 & 0 & 0 & 0 & 0 & 0 & 0     \\ 0 & 0 & 0
& 0 & 0 & 0 & 0         \\ 1 & 0 & 0 & 0 & 0 & 0 & 0       \\
\end{tabular}
\right ] V_{1}~~~~~~~~~~~~~~~~~~~~~    \nonumber    \\
\nonumber    \\
<\overline{\Phi}> = <\Phi> =[0 ~0 ~0 ~0 ~0 ~0 ~1]
V_{2}~~~~~~~~~~~~~~~~~~~~~~~~~~
\nonumber \\
\nonumber    \\
< {S_{1}}^{A} >=\biggl(\frac{{M_3}^2 ~M_{4}}{h_{2}^2
h_{1}} \biggl)^\frac{1}{3}~~~~~~~~
< {S^{1}}_{A} >=\biggl(\frac{M_{4}^{2}~M_{3}}{h_{1}^2~
h_2}\biggr)^\frac{1}{3}~~~~~~~~~~~~~~~~~ \nonumber \\
\nonumber    \\
< {S_{2}}^{B} >=\biggl(\frac{{M_4}^2 ~M_{2}}{h_{2}^2
h_{1}} \biggl)^\frac{1}{3}~~~~~~~~
< {S^{2}}_{B} >=\biggl(\frac{M_{2}^{2}~M_{4}}{h_{1}^2~
h_2}\biggr)^\frac{1}{3}~~~~~~~~~~~~~~~~~ \nonumber \\
\nonumber    \\
< {S_{3}}^{C} >=\biggl(\frac{{M_2}^2 ~M_{3}}{h_{2}^2
h_{1}} \biggl)^\frac{1}{3}~~~~~~~~
< {S^{3}}_{C} >=\biggl(\frac{M_{3}^{2}~M_{2}}{h_{1}^2~
h_2}\biggr)^\frac{1}{3}~~~~~~~~~~~~~~~~~
\end{eqnarray}

where

\begin{center}
$V_{1}^{2}=~\frac{M_{1}}{\lambda_{1}}(6\frac{m_{1}}{\lambda_{1}} + 2
\frac{m_{2}~h}{\lambda_{1}~\lambda_{2}})+
5\frac{m_{1}~m_{2}}{\lambda_{1}~\lambda_{2}}+6\frac{{m_1}^{2}}{h~\lambda_{1}}+
\frac{h}{\lambda_{2}}(\frac{m_{2}}{\lambda_{1}})^2$  \\
\vspace{3mm}

$V_{2}^{2}=(\frac{{M_1}}{\lambda_{1}})^{2} \frac{h}{\lambda_{2}}+
\frac{M_{1}}{\lambda_{1}}(4\frac{m_{1}}{\lambda_{2}}+\frac{m_{2}}{\lambda_{2}}
\frac{h}{\lambda_{2}})+6\frac{m_{1}m_{2}}{\lambda_{2}~\lambda_{2}}+
\frac{{m_1}^2}{h~\lambda_{2}}$ \\
\end{center}

 In this vacuum the hierarchy of the symmetry
breakings is the following: at the scale $M_{GUT}$ the gauge $SU(7)$
symmetry is broken down to $SU(3)_{C}\otimes SU(3)_{W}\otimes
U_{1}(1)$ by the $\Sigma$ and $(\overline{\Phi}, \Phi)$ which develop the
 large $(\sim M_{GUT})$ VEVs, at the ``geometrical scale'' $(\sim
\sqrt{M_{GUT}~m_{3/2}})$ the VEVs  $<H^{A}> =< \overline {H}_{A}>$
break $SU(3)_{W}\otimes U_{1}(1)$ down to $SU(2)_{W}\otimes
U(1)_{Y}$.  So, according to a general argument of Sec.2
 in this case we have the two pairs of
light Higgs doublets.

 According to the
standard scenario of $SU(2)_W\otimes U(1)_Y$ breaking by radiative
corrections [8] the main negative contribution to the one of the light
Higgs $[mass]^2$ comes from the loops with the matter fermions, corrections by
the top quark exchange (due to its largest Yukawa coupling constant).The
other ones do not have coupling with light matter fermions 
 (as it is shown in section 4) So the flavour changing neutral
processes are avoided.

\section {Matter field sector }

 We place each generation of quark~-~lepton superfields we place in the set of
anomaly free set of $SU(7)$ representations. These
are two $\overline {21}$, $\overline {21'}$  and three $35$
representations which we denote as $\overline
{21}$ and $35^{A} (A = 1, 2, 3$ and is the indexe of
 $SU(3)_{cust}$ symmetry) respectively.  Their decompositions under the
$SU(5)\otimes SU(2) \subset SU(7)$ subgroup are
\begin{center}
$\overline {21} =(\overline {10},1)+(\overline{5},2)+(1,1)$

$35^{A}=(\overline {10},1)^{A}+(10,2)^{A}+(5,1)^{A}$
\end{center}

 The masses of the quarks and leptons are generated by the following
``Yukawa'' couplings:

 a)$f_{1} \cdot \overline {21} \cdot35^{A}\cdot \overline{H_{A}}-$
for the ``down'' type quarks and leptons

b)~ $f_{2} \cdot 35^{A}\cdot 35^{B} \cdot  H^{C} \cdot\epsilon_{ABC}-$
for the ``up'' type quarks

For the heavy multiplets we have the following ``Yukawa'' couplings.

 c)~ $f_{3}\frac{1}{M_{GUT}}  \cdot \overline{21} \cdot 35^{A} \cdot
 \overline{\Phi} \cdot {S^{i}}_{A}$

 d)~ $f_{4}\frac{1}{M_{GUT}}  \cdot {35}^{A} \cdot {35}^{B}  \cdot
  \Phi \cdot {S_{i}}^{C} \cdot \epsilon_{ABC}$

 e)~$f_{5} \cdot 35^{A}\cdot 35^{B}\cdot H^{C}\epsilon_{ABC}$

 f)~$f_{6} \cdot \overline{21} \cdot {35}^{A} \cdot H_{A}$

(c) and (d) are nonrenormalizable coupling and have to be understood be
considered as 
effective operators. For example, this term can be easily generated trough
the
heavy (with mass $\sim  M_{GUT}$) particle exchange [10], namely 
exchange of the representations of:
 $ ({35}^{AB} +\overline{35}_{AB})$ and
$({21}^{A} +\overline{21}_{A})$ can do the job.  The
relevant couplings for the case (d) are
\begin{center}
$\kappa_{1}{35}^{A}\cdot {35}^{BC}\Phi\epsilon_{ABC}+ M_{GUT}
\overline{35}_{BC}\cdot 35^{BC}+ \kappa_{2}\overline{35}_{BC}\cdot
{35}^{B}\cdot {S_{i}}^{C}$
\end{center}
for the case (c)
\begin{center}
$\kappa_{2}{35}^{A}\cdot \overline{21}_{A}\overline{\Phi}+
M_{GUT} \overline{21}_{A}\cdot {21}^{A}+
\kappa_{4}\overline{21} \cdot {21}^{A}\cdot {S^{i}}_{A}$
 \end{center}
It is assumed that 
 $\kappa_{1}$, $\kappa_{2}$, $\kappa_{3}$ and
$\kappa_{4}$ are of order $\sim 1$

From the (c) and  (d) couplings, the submultiplets:  three  pair of
$(\overline {10},1) + (10.1)$ and one pair $(\overline 5,1) +
(5,1)$ acquire masses of order $M_{GUT}$. From the coupling (e) and
(f) a single pair $(\overline{5},1)\cdot (5,1)$ acquire masses of order $\sim
\sqrt{m_{3/2} M_{GUT}}$.

Thus, we have per family: $(5,1)+(\overline{5} ,1)$
submultiplets
at the ``geometrical scale'', and three pairs of the $(10,1)+(\overline
{10},1)$ and one pair of the $(5,1)+(\overline{5} ,1)$ at the GUT scale.

\section {Vector-like fermionic superfields}

Besides the problem of the masses for quarks and leptons  which could be
solved in the above mentioned way there is  a problem of
unification. Because  extension of the model with  the intermediate
scale the unification of the gauge couplings in general may be spoiled.
So, we consider extension of the model, as to include not only chiral
superfields but also vector-like fermionic superfields (VLFS).

Imagine that we have all possible antisymmetric and fundamental
vector-like
fermionic representations
\begin{center}
$\overline{7}+7$ ;    ~~~ $\overline{21}+21$ ;
~~~$\overline{35}+35$
\end{center}
The following terms are possible in the ``Yukawa'' superpotential :
 \begin{eqnarray}
W_{VLFS}=m_{7} \overline{7}\cdot 7 + \eta_{1}
\overline{7} \cdot \Sigma \cdot 7 + m_{21} \overline{21} \cdot 21 +
\eta_{2} \overline{21} \cdot \Sigma \cdot 21 + \eta_{3}
\overline{35}\cdot \Sigma\cdot 35~~~~~  \nonumber \\
~~~~~+m_{35} \overline{35}\cdot 35
+\chi_{1} \overline
{21} \cdot 35 \cdot \overline {\Phi} + \chi_{2} \overline {35} \cdot
21\cdot \Phi + \chi_{3} \overline{21} \cdot 7\cdot \Phi+ \chi_{4}
21\cdot \overline{7}\cdot \overline{\Phi}~~~~~
\end{eqnarray}
where $\eta_{1}$, $\eta_{2}$, $\eta_{3}$, $\chi_{1}$, $\chi_{2}$,
$\chi_{3}$, $\chi_{4}$ are dimensionless constants and $m_{7}$,
$m_{21}$, $m_{35}$ are mass parameters of order $m_{3/2}$. 
As was shown $\Sigma$s  field
VEV has the form:
\begin{center}
$<\Sigma>=\frac{M_{GUT}}{\lambda_1}diag[1~ 1~ 1~ -1~ -1~ -1~ 0] +
O(m_{3/2})$
\end{center}
it is easy to check that $\eta$ terms can not 
generate masses of the following fragments 
\begin{center}
$(\overline 3,\overline 3) +(3,3)~~~~~~ from
~~~~~~~(\overline{35}+35)$

$(\overline 3,\overline3) +(3,3) ~~~~~~~from
~~~~~~~(\overline{21}+21)$
\end{center}
(these are submultiplets under the $SU(3)_{W}\otimes SU(3)_{C}$ subgroup
of $SU(7)$) the  others acquire masses of order $\sim M_{GUT}$. The
multiplets $(\overline {7}+7)$ get mass of order $\sim M_{GUT}$.  This
mechanism is known as the split-multiplet mechanism and was proposed in
ref [9]. From the $\chi$ terms the split fragments get masses $\sim
{\chi}_{i} \cdot M_{GUT}$ and if we suppose that $\chi_{i}$ are in the
$0.01\div 0.007$ interval we have the split fragments in $M_{SPM} \sim
1.5\cdot 10^{15}~\div ~~5\cdot 10^{15}$ GeV region which lead to the
successful unification (see table 1) and this scale we denote as
$M_{SPM}$

\section{\bf Gauge coupling unification}

Now, let us begin the renormalization group (RG) analysis of our model.
The two-loop RG equations for the running gauge couplings of general
effective $G_{1}\otimes G_{2}\otimes ...$ gauge theory have the well known
form \begin{equation} \mu\frac{d}{d\mu}\alpha^{-1}_{i}=-\frac{1}{2\pi}
\biggl(b_i+\frac{b_{ij}}{4\pi }\alpha_j+O(\alpha^{2}_i)\biggr)
\label{}
\end{equation}
where $\alpha _{i}(\mu)$ is the running gauge constant corresponding to
$G_{i}$ group, while $b_{i}$  and $b_{ij}$  are one and two-loop b-factors
[11] respectively.

There are three energy regions in our case $M_{Z} - M_{SUSY}
(M_{SUSY}\sim m_{3/2})$, $M_{SUSY} - M_{I}$, and $M_{I} - M_{G}$  where the
b-factors are:

1)~$M_{Z} - M_{SUSY}$, SM region
\begin{eqnarray}
{b_i^{SM}}=
\left( \begin{array}{ccc}
 4 & -\frac{10}{3} & -7
\end{array} \right) + N_{H}\cdot \left( \begin{array}{ccc}
 \frac{1}{10} & \frac{1}{6} & 0
\end{array} \right)   \nonumber \\
{b_{ij}^{SM}}=
\left( \begin{array}{ccc}
\frac{19}{5} & \frac{9}{5}  & \frac{44}{5} \\
\frac{3}{5} &  \frac{11}{3} & 12 \\
\frac{11}{10}  & \frac{9}{2} & -26
\end{array} \right)  +
N_{H}\cdot \left( \begin{array}{ccc}
\frac{9}{50}  & \frac{9}{10} & 0 \\
\frac{3}{10} &  \frac{13}{4} & 0 \\
0 & 0 & 0
\end{array} \right)
\end{eqnarray}
Here $N_{H}$ is the number of light Higgs doublets;

2)~$M_{SUSY} - M_{I}$, SUSY SM region
\begin{eqnarray}
{b_i^{SSM}}=
\left( \begin{array}{ccc}
 6 & 0 & -3
\end{array} \right) + N_{H}\cdot \left( \begin{array}{ccc}
 \frac{3}{10} & \frac{1}{2} & 0
\end{array} \right)   \nonumber \\
{b_{ij}^{SSM}}=
\left( \begin{array}{ccc}
\frac{38}{5} & \frac{18}{5} & \frac{88}{3} \\
\frac{6}{5} &  18 & 24 \\
\frac{11}{5}  & 9 & 14
\end{array} \right)  +
N_{H}\cdot \left( \begin{array}{ccc}
\frac{9}{50} & \frac{9}{10} & 0 \\
\frac{3}{10} &  \frac{7}{2} & 0 \\
0 & 0 & 0
\end{array} \right)
\end{eqnarray}
3)~$M_{I} - M_{G}$ region
\begin{eqnarray}
{b_i^{G_{I}}}=
\left( \begin{array}{ccc}
 9 & 0 & 0
\end{array} \right) +
N_{H}\cdot \left( \begin{array}{ccc}
\frac{1}{4} & \frac{1}{2} & 0
\end{array} \right)
\nonumber\\
{b_{ij}^{G_{I}}}=
\left( \begin{array}{ccc}
 9 & 24 & 24 \\
 3 & 48 & 24 \\
 3 & 24 & 48
\end{array} \right) +
N_{H}\cdot \left( \begin{array}{ccc}
 \frac{1}{12} & \frac{4}{3} &  0  \\
 \frac{1}{6}  & \frac{17}{3} & 0 \\
 0 & 0 & 0
\end{array} \right)
\end{eqnarray}

In addition we have to compute the contributions of split-fragments from
$(\overline{21}+21)$ and $(\overline{35}+35)$
\begin{eqnarray}
{b_i^{SPM}}=
\left( \begin{array}{ccc}
 0 & 6 & 6
\end{array} \right),   \qquad {b_{ij}^{SPM}}=
\left( \begin{array}{ccc}
 0 & 0 & 0 \\
 0 & 68 & 32 \\
 0 & 32 & 68
\end{array} \right)
\end{eqnarray}
We assume that they begin to play the role only from the energy scale
$M_{SPM}$, where $M_{SPM}$ is a free parameter which fixed from the
unification condition:
\begin{equation} \alpha _{G}(M_G)=\alpha
_{1}(M_G)=\alpha _{2}(M_G)=\alpha _{3}(M_{GUT}) \label{} \end{equation}

and the value of this scale we can explain from (9). The gauge
couplings $\alpha _{3}(\mu), \alpha _{2}(\mu), \alpha _{1}(\mu)$
correspond to $SU(3)_c, SU(3)_W, U(1)$ gauge groups (in the $M_{I} <
\mu < M_{G}$ region) respectively.  At the  scale $M_{I}$ they are
related to $\alpha _{c}(\mu), \alpha _{W}(\mu)$, and $\alpha
_{Y}(\mu)$ gauge couplings which correspond to $SU(3)_s, SU(2)_W,
U(1)_{Y}$ groups (in the $M_{Z} < \mu < M_{I}$ region ) respectively,
by equations:  \begin{eqnarray}
\alpha_{s}(M_{I})=\alpha_{3}(M_{I}),~~~
\alpha_{w}(M_{I})=\alpha_{2}(M_{I}), \nonumber  \\
\alpha_{Y}^{-1}(M_{I})=\frac{4}{5}\alpha_{1}^{-1}(M_{I})+
\frac{1}{5}\alpha_{2}^{-1}(M_{I})
\label{}
\end{eqnarray}

We have solved equations (13) numerically with b-factors (14), (15), (16),
(17) and conditions (18), (19) using as input parameters [12]
\begin{eqnarray}
\alpha _{s}=0.117\pm 0.005        \nonumber \\
sin^{2}\theta _{W}=0.2319\pm 0.0005 \nonumber \\ \alpha^{-1}_{EM}=127.9\pm
0.02 \label{}
\end{eqnarray}
The result of computations for the low
values of (17) for $\alpha_{s}$  end low values of $M_{SUSY}=250 GeV$
is presented in fig. 1. We have also
plotted the flow of running gauge coupling constants in Table 1

Note that the grand unification scale is  close to $M_{Pl}$.
Such a large unification scale avoids the Standard SUSY GUT troubles
with $d=5$ operator induced baryon decay, since the Higgsino mass in our
case is of the order of $M_{H_{c}}\sim M_{G(SU(7))}\sim 10^{18}$ GeV
(whereas in the standard SUSY GUT it is $\sim 10^{16}$ GeV), so the proton
lifetime is increased relatively to the standard SUSY GUT case by the
factor $\sim \frac{M^{2}_{G(SU(7))}}{M^{2}_{G(SU(5))}}\sim 10^{3\div 4}$ and
no constraints on the SUSY parameter space are required.

\section{\bf Conclusions}

We have studied the $SU(7)$ SUSY GUT with the ``custodial symmetry''
mechanism for the explanation of the DT hierarchy, which naturally leads
to existence of the intermediate $G_{I}$ symmetry scale $M_{I}$ in the
desert between $M_{SUSY}$ and $M_{G}$. 

To obtain the gauge coupling
unification we have introduced an additional pair of light Higgs doublets
and split the vector-like matter superfields. 

As it is shown in this model it is possible to get unification of the 
gauge coupling constants for all the
reasonable values of $\alpha_{s}$ and $M_{SUSY}$ and a correct value of 
$sin^2 {\theta}_{W}$. 

Since the unification 
scale is close to $M_{Pl}$, there is no problem with $d=5$ operator
induced baryon decay. 

On the other hand, the introduction of four
light doublets give the chance to obtain the correct value not only
for $m_{b}/m_{\tau}$ but also for $m_{s}/m_{\mu}$ in the manner of
ref.[13].


\section{Acknowledgements}

It is a great pleasure for me to thank 
 H.~Asatrian, Z.~Berezhiani, J.L.~Chkareuli, G.~Dvali, 
A.~Kobakhidze, L.~Rurua, E.~Sarkisyan, 
A.~Smirnov, and Z.~Tavartkiladze for useful comments and help during the
preparation of this paper. 

This work was supported in part by International Science Foundation (ISF)
under the grant $No.\rm MXL000$ and $No.\rm MXL200$.

\newpage

{\bf Figure Captions}

\vspace{1 cm}

Fig.1: Gauge coupling unification with two-loop evolution for 
$M_{SUSY}=$250 GeV, 

\hspace{1.2cm} ${\alpha}_s=0.112$, $sin^2{\theta}_W=0.2314$

\newpage
\vspace*{5cm}
{\bf Table~1}
\medskip
\begin{center}
\begin {tabular}{|c|c|c|c|c|c|}
\hline
& & & & &        \\
$\alpha_{s}$ &  $M_{SUSY}, GeV$ & $M_{I}, GeV$ & $M_{SPM}, GeV$ &
$M_{G}, GeV$ & $\alpha ^{-1}_{G}$ \\
& & & & & \\ \hline
& & & & & \\
0.112 & $2.5\cdot 10^2$ & $9.2\cdot 10^{9}$ & $4.47\cdot
10^{15}$ & $3.3\cdot 10^{17}$ & 12.7 \\
& & & & & \\ \hline
& & & & & \\
0.117 & $2.5\cdot10^2$ & $1.27\cdot 10^{10}$ & $4.26\cdot
10^{15}$ & $6.46\cdot 10^{17}$ & 11.66\\
& & & & & \\ \hline
& & & & & \\
0.122 & $2.5\cdot10^2$ & $1.72\cdot10^{10}$ & $4.17\cdot10^{15}$
& $1.17\cdot 10^{18}$ & 10.73 \\
& & & & & \\ \hline
& & & & & \\
0.112 & $10^3$ & $1.55\cdot 10^{10}$ & $3.09\cdot
10^{15}$ & $4.57\cdot 10^{17}$ & 14\\ & & & & & \\ \hline
& & & & &        \\
0.117 & $10^3$ & $2.14\cdot 10^{10}$ & $2.95\cdot 10^{15}$ &
$4.57\cdot 10^{17}$ & 13 \\
& & & & & \\ \hline
& & & & & \\
0.122 & $10^3$ & $2.85\cdot 10^{10}$ & $2.88\cdot
10^{15}$ & $8.13\cdot 10^{17}$ & 12.1 \\
& & & & & \\ \hline
\end{tabular}
\end{center}

\newpage

\epsfbox{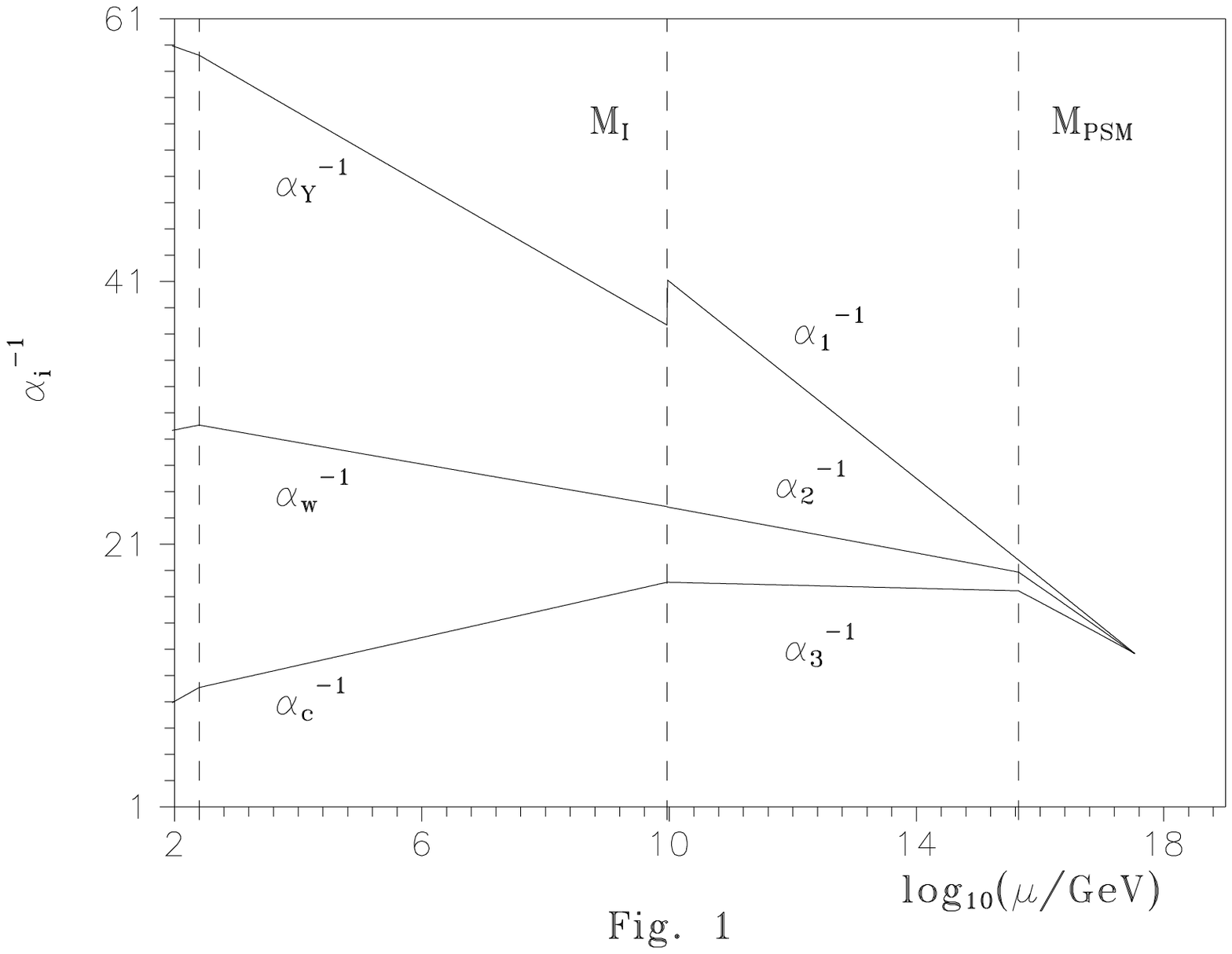}

\end{document}